# AUTOMATED DATA EXTRACTION OF BAR CHART RASTER IMAGES

*Andrew W. Browne, 850 Health Sciences Road, Irvine, CA 92697, USA; abrowne1@uci.edu; 949-824-6256

Alex Carderas[1], Ye Yuan[1], Itamar Livnat[2], Ryan Yanagihara[3], Rosita Saul[4], Gabrielle Montes De Oca[4], Kai Zheng[5], Andrew W. Browne[2]

[1]Department of Computer Science, Donald Bren School of Information and Computer Sciences, University of California, Irvine, Irvine, California, United States, [2]Department of Ophthalmology, Gavin Herbert Eye Institute, University of California, Irvine, Irvine, California, United States, [3]Department of Pediatrics, John A. Burns School of Medicine, University of Hawaii at Manoa, Honolulu, Hawaii, United States, [4]Department of Biological Sciences, Schmid College of Science and Technology, Chapman University, Orange, California, United States, [5]Department of Informatics, Donald Bren School of Information and Computer Sciences, University of California, Irvine, Irvine, California, United States,

## Abstract

Objective: To develop software utilizing optical character recognition toward the automatic extraction of data from bar charts for meta-analysis.

Methods: We utilized a multistep data extraction approach that included figure extraction, text detection, and image disassembly. PubMed Central papers that were processed in this manner included clinical trials regarding macular degeneration, a disease causing blindness with a heavy disease burden and many clinical trials. Bar chart characteristics were extracted in both an automated and manual fashion. These two approaches were then compared for accuracy. These characteristics were then compared using a Bland-Altman analysis.

Results: Based on Bland-Altman analysis, 91.8% of data points were within the limits of agreement. By comparing our automated data extraction with manual data extraction, automated data extraction yielded the following accuracies: X-axis labels 79.5%, Y-tick values 88.6%, Y-axis label 88.6%, Bar value <5% error 88.0%.

Discussion: Based on our analysis, we achieved an agreement between automated data extraction and manual data extraction. A major source of error was the incorrect delineation of 7s as 2s by optical character recognition library. We also would benefit from adding redundancy checks in the form of a deep neural network to boost our bar detection accuracy. Further refinements to this method are justified to extract tabulated and line graph data to facilitate automated data gathering for meta-analysis.

## Introduction

Academic research publication rates are accelerating. In 2018 alone, more than 2.5 million new scientific and technical papers were published.[1] These papers form a network of information, each a piece of a puzzle that, when arranged, reveals new insights into the natural world. For any individual, manually compiling a subset of this data for meta-analysis is an endless uphill battle.

Traditional approaches to automating this task focus solely on extracting data from text and often neglect images and figures. In biomedical research, a figure often provides as much information as a normal abstract.[2] Because of this, an automated meta-analysis system capable of extracting information from figures is needed. Such a system can be assembled using deep learning models and simple computer vision techniques. In this paper, we will discuss a pipeline for extracting information from bar chart figures.

While existing pipelines aim to automate data extraction from images, most address the problems of image classification, text retrieval, and image retrieval.[3-5] Little work has successfully automated the task of retrieving data from bar charts. Huang et al. proposed a system for understanding imaged infographics of bar and pie charts.[6] The system architecture consists of 3 modules: graphics recognition, text recognition, and understanding. They achieved a graphics recognition accuracy of 93.75% and a text block classification accuracy of 88.22% on bar chart images. He et al. describes a workflow for bar chart image mining.[7] They propose a system comprised of 6 components: figure extraction, image preprocessing, bar segment detection, in-image text recognition, panel segmentation, and quantitative information extraction. Using hand-coded rules and a convolutional neural network (CNN), 39.81% of bar panels were successfully extracted, 82.40% thereof were quantitatively correct. Liu et al. introduce relational networks (RNs) as a deep learning solution to extract data from bar and pie charts.[2] Matched RN components, such as bars and axes, are used to predict coordinate-value relations. The Faster R-CNN model is used to locate feature objects such as text and axes.[8] Over a diverse set of figures, they achieved an error of less than 1% on 64.75% of images. Ultimately, each of these approaches extracted data from bar charts with high accuracy, but the challenge of extracting information over diverse bar chart representations necessitates additional approaches toward the goal of automated bar chart analysis in all scientific papers.

**Meta-Analysis**

In medical research, many individual studies lack the statistical power and even reproducibility to provide conclusive evidence for changing clinical practice. For this purpose, meta-analyses are often employed to synthesize the results gathered by multiple research papers toward giving generalizable guidelines that can become the standard of care. A meta-analysis accomplishes this goal objectively by increasing statistical power compared to any single publication.[9] In general, a meta-analysis begins with gathering relevant literature through comprehensive database searches and then extracting and combining the data from those papers. While online database searches facilitate literature searches, extracting information from the literature is a time-consuming task as publications lack standard representations of results necessitating extensive manual labor to sort through volumes and identify relevant publications. We are interested in developing an algorithm that can automate data extraction from published papers, beginning with figure data on visual acuity in age-related macular degeneration (AMD).

AMD is a progressive retinal degenerative disease and a leading cause of blindness in the elderly in developed countries.[10-14] AMD is divided into two categories: the dry form and the wet form, the latter of which is termed neovascular AMD.[15] Neovascular AMD treatments include laser-based treatments, and intraocular injections with anti-vascular endothelial growth factor (anti-VEGF) medications including bevacizumab, ranibizumab, aflibercept and brolucizumab.[16] Ultimately, the goal of these treatments is to control the progression of diseases and take steps towards improved visual acuity (VA) outcomes. Often, these treatments will also improve central foveal thickness (CFT) as a result of reducing neovascularization. With relatively few outcomes of interest, many randomized controlled trials (RCTs) addressing AMD, and due to the burden of AMD, we chose RCTs related to treatments for AMD as our first test of automated bar chart data extraction.

The objective of this paper is to describe a data extraction pipeline to parse figure data from documents related to AMD. The processed data can then be used to observe the changes in VA and CFT resulting from different treatments. While AMD is the prototypical disease used as an example in this work, we expect this to be expandable to all disease states across disciplines in medicine. Thus, we propose a data extraction pipeline to automatically extract bar chart data from documents containing raster images.

## Methods

Our data extraction pipeline consists of 4 steps: figure extraction, text detection, image disassembly, and data extraction (Figure 1). During figure extraction, the algorithms scrub figures and images from a document and categorize them by their figure type (bar graph vs line graph vs other). Text recognition preprocesses each figure and performs optical character recognition (OCR). Features like bars, axes, and legends are identified during image disassembly. Finally, the figures are reconstructed, pairing text with feature objects in the data extraction step and data is tabulated.

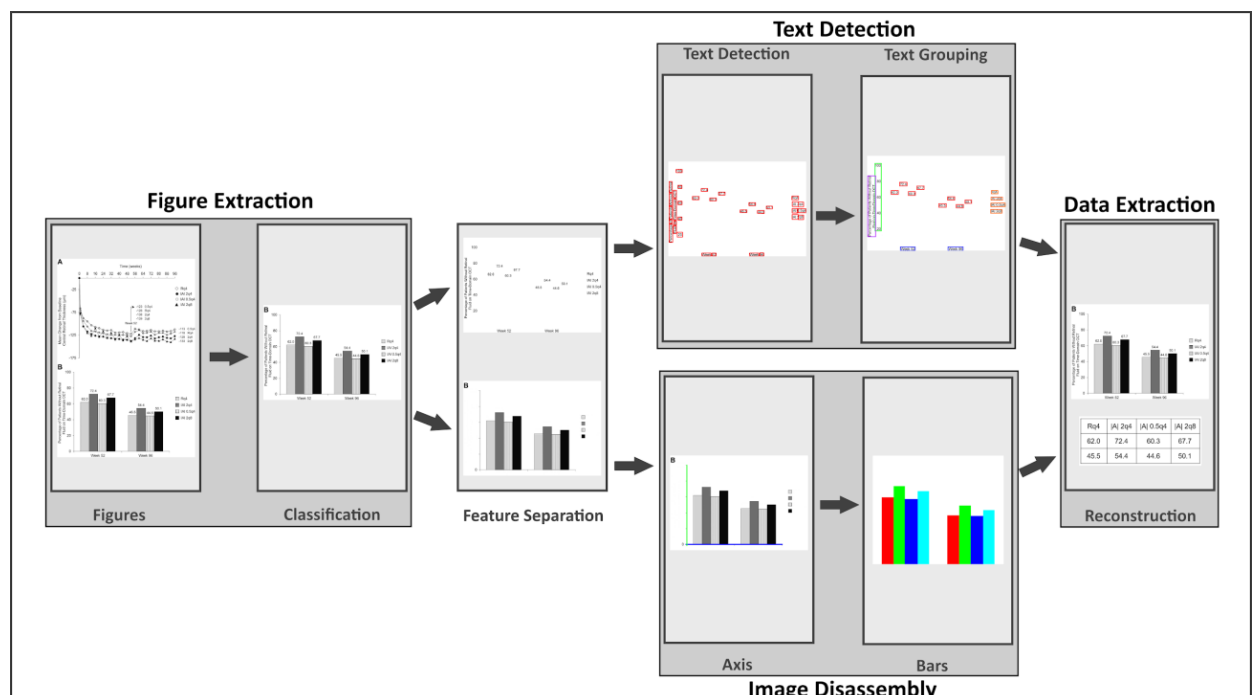

Figure 1. Data extraction pipeline.

***Figure extraction.*** The pipeline begins by finding all figures stored inside a document. Images are extracted using MuPDF and segmented by their constituent panels.[17-18] Image segmentation is indispensable when extracting information from biomedical images, as many images are composed of multi-panel figures. Constituent panel segmentation is preferred to gap segmentation because it addresses the issues of under and over segmenting. This can be achieved by labeling and grouping pixels by their intensity and calculating the smallest bounding box that encapsulates each set of labels. Each of the extracted panels are passed through an Inception v3 architecture model trained on ImageNet images with a new top layer trained to recognize bar charts.[19]

***Text detection.*** To improve the accuracy of OCR, each figure is preprocessed to remove noise and non-text features associated with bars and axes. Otsu's binarization is applied to the images to increase contrast, and contours are found using simple border following.[20] Contours with an area outside one standard deviation of the average are rejected. Contours with a non-zero pixel average of less than 25% are also rejected. These two steps remove large bars and thin interconnected lines from the image. What remains is a list of contours that represent bounding boxes of text. Using these bounding boxes, a text mask of the original image is created. Finally a 2x upscale of the mask is applied using Waifu2x, a super-resolution deep convolutional neural network(SRCNN).[21] Results from ICDAR2015 show that SRCNN's outperform state of the art techniques such as the Zeyde and A+ dictionaries in both reconstruction measures and OCR accuracy.[22-24] Finally, text is extracted from the upscaled image using Tesseract-OCR.[25]

*Image disassembly.* To understand a figure as a whole, we must first break it down into its components. The first key component to extract is the figure's axes, as its location will help shed light on where bars will be located. Before extracting the axes, we can improve the accuracy by removing text from the image. This can be achieved by subtracting the image mask obtained in the text detection step from the original image. Next, a Gaussian blur with a kernel size of 5 and an adaptive threshold using the integral image is applied.[26] The axes of the graph are found using the Canny edge detection and probabilistic Hough transform algorithms.[27-28] The two largest lines that create a 90-degree corner are chosen as the axes. The image is then cropped using the rectangular region formed by the axes.

A morphological opening operation with a kernel size of 5 is applied to the image to remove gridlines. Contour corners are extracted using simple chain approximation and paired to form vertical lines. Two corners form a vertical line if the difference between their x-coordinates is zero. The two nearest vertical lines sharing a vertex with the same y-value are grouped to form a bar. The pairing process requires a starting corner to lie on the x-axis and ends when a subsequent paired corner touches the x-axis again. Small slices from the center of individual bars are taken and used for template and color matching. Bars with the same color and template are grouped.

*Data extraction.* The figure's title is obtained from the text located closest to the top-middle of the figure. Text located perpendicular to the y-axis are classified as y-tick values and text parallel to the x-axis are x-tick values. Subsequent text positioned to the left of y-tick values and below x-tick values are used as labels for their respective axes. Quantitative data related to bars can be obtained through two means: text located directly above individual bars, and the coordinate-text relation of bar height and y-tick values. When available, text above individual bars is used as the preferred method of obtaining quantitative data as it is the most accurate. However, figures do not always present this information and their values must be approximated. To approximate bar values, the average distance between y-tick text is calculated and used to determine a coordinate-value relationship. This approximation technique assumes that y-tick values scale linearly and will not work on figures with a logarithmic scale. Figure 2 illustrates an example of an annotated bar chart output by our software.

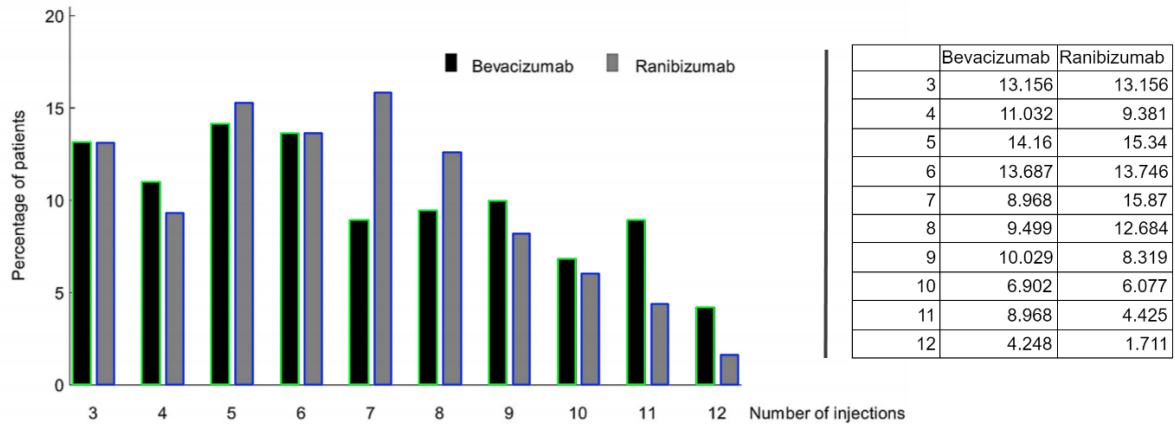

*Figure 2. Automatic Data Extraction Using Our Pipeline*

***Randomized controlled trials.*** To test our pipeline we compiled a corpus of 10 RCTs related to AMD from PubMed Central with significant statistical importance to VA and CFT analysis.[29-38] Of these documents, a total of 58 figures were extracted, 48% of them (28) were bar charts. Data was manually extracted by two independent graders from the figures using WebPlotDigitizer.[39] The same figures were then passed through our automated data extraction pipeline for comparative analysis. While many RCTs are published for AMD, the represented RCTs were chosen on the basis of containing bar chart data. In addition, these were recent RCTs in the era of evolving treatment of AMD. An example of WebPlotDigitizer manual data extraction from the same bar chart as Figure 2 is illustrated in Figure 3.

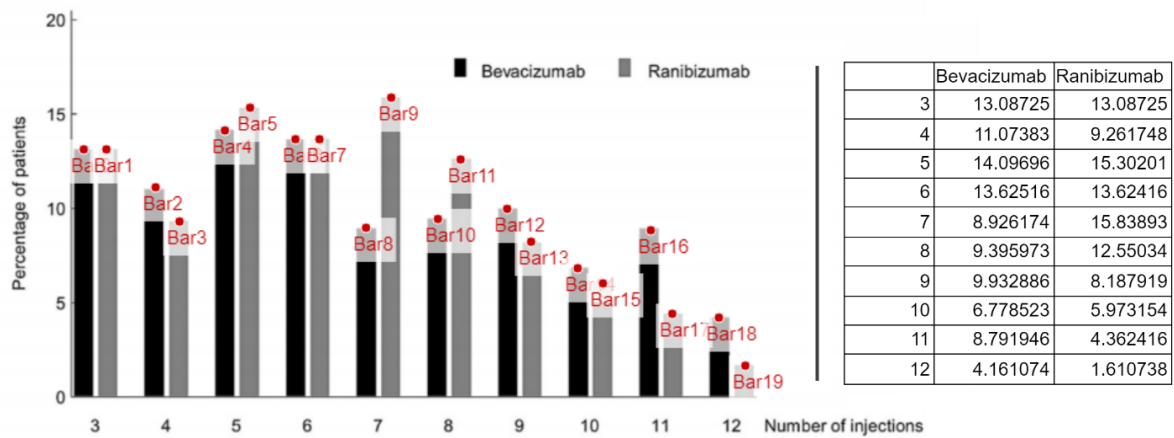

*Figure 3. Manual Data Extraction Using WebPlotDigitizer.*

***Data Analysis.*** To investigate the agreement between data, obtained through manual and automated extraction, Bland-Altman analysis was employed on 292 data points extracted from the bar chart figures. This number excludes 77 data points that were not were not detected during the extraction process. Bland-Altman analysis evaluates the differences between two measurement methods by identifying differences between the mean of two measurements (bias) and each of the two measurements, and by estimating an agreement interval defined as 2 standard deviations from the bias.[40] While this analysis does not say if the agreement is sufficient, it does quantify a range of agreement within which 95% of differences should lie.

For the purposes of statistical analysis we measured change in VA as the percent of patients who lost less than 15 letters on an ETDRS visual acuity measurement chart. An ETDRS chart is used as a standard testing method for determining VA in patients suffering from AMD and other visually impairing conditions.[41] Our analysis evaluates the change in VA of patients undergoing one of three different treatments: Bevacizumab 1.25 mg monthly, Ranibizumab 0.5 mg monthly, and Ranibizumab 0.5 mg treat-and-extend (T&E).

| OBJECT | ACCURACY |
|---|---|
| X-TICK VALUE | 85.2 |
| X-AXIS LABEL | 79.5 |
| Y-TICK VALUE | 81.4 |
| Y-AXIS LABEL | 88.6 |
| BAR VALUE (<1% ERR) | 46.6 |
| BAR VALUE (<2% ERR) | 67.5 |
| BAR VALUE (<5% ERR) | 88.0 |

*Table 1. Data extraction accuracy.*

## Results

Of the 28 bar charts extracted from the documents, 92.8% of them were successfully recognized by the CNN. Additionally, 96.0% of text blocks were correctly identified and 79.1% of all bars were detected. As shown in Table 1, each feature object and text object have an accuracy score that represents the percentage of items that matched exactly to the manually extracted data. These scores reflect only those data points which were successfully detected by the pipeline. For labels and tick values, an exact match means that the automatically extracted string is identical to the manually extracted one. For bar values, the accuracy score represents the percentage of values that were within 1%, 2%, and 5% of the manually extracted data.

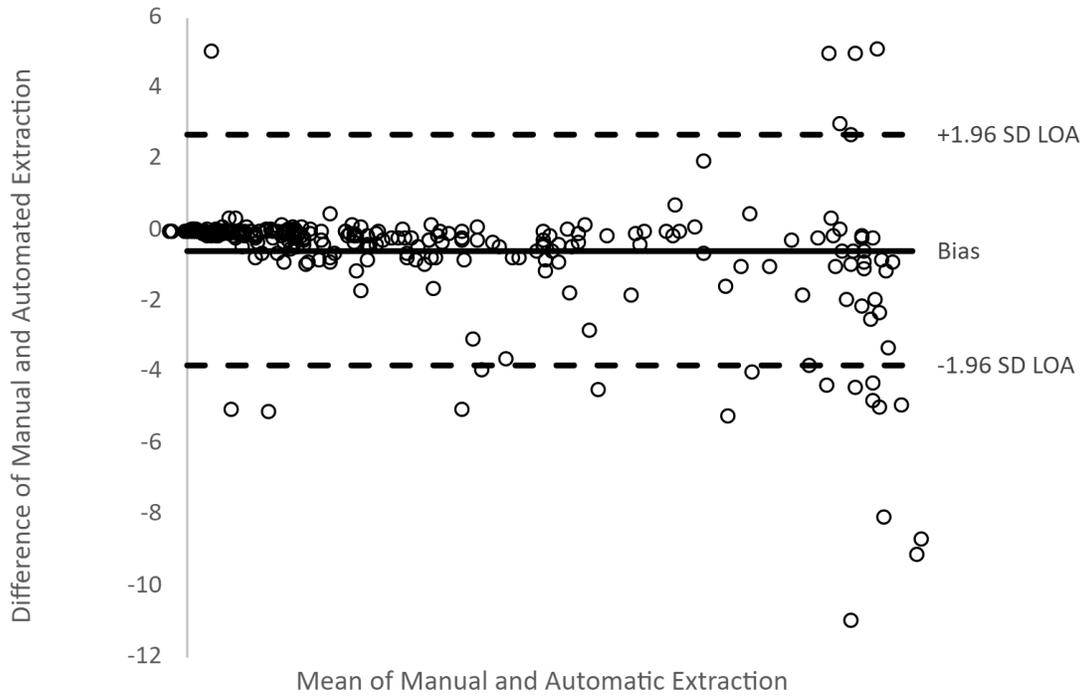

*Figure 4. Bland-Altman Analysis.*

Results from the Bland-Altman analysis in Figure 4 show that with a bias of -0.544 and a standard deviation of 1.652, 91.8% of data points (268 out of 292) lie within the limits of agreement. A side by side comparison of manually and automatically extracted data is shown in Figure 5. These scatterplots reveal the percent of patients that lost less than 15 letters from an ETDRS chart throughout a 12 month period.

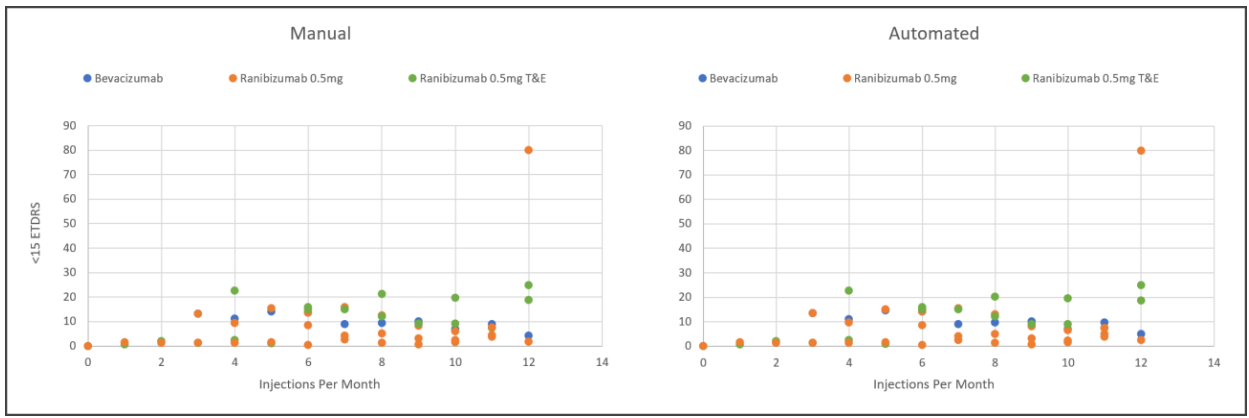

*Figure 5. Change in ETDRS Letters Manual vs Automated*

## Discussion

Manually compiling data from publications for meta-analysis is a considerable challenge given the increased rate at which publications are being produced. Multiple software approaches have been undertaken to automate data extraction , but none of these approaches have yielded a high enough degree of accuracy for routine use to reduce the time to gather data for a meta-analysis. We developed software with a multi-step heuristic approach utilizing OCR and a CNN toward the goal of automated data extraction for meta-analysis, beginning with publications pertaining to AMD, a progressive retinal degeneration with heavy disease burden in the field of ophthalmology.

When comparing our results to other data extraction pipelines, we achieved a figure recognition score of 91.2% vs 93.2% and a text block detection accuracy of 96.0% vs 88.22% compared to Huang et al. Our bar panel detection rate of 79.1% showed considerable improvements compared to the 39.81% detection rate by He et al. Finally, our < 1% error rate of extracted data points achieved a 46.5% accuracy compared to 64.75% of Liu et al. We believe these lower accuracy scores are reflective of the static nature of hand coded rules. Liu et al's. use of relational networks and machine learning for data extraction is a testament to the ability of neural networks to perform well on new data. Our software model would benefit from adding redundancy checks in the form of deep neural networks to aid our heuristic approach.

The Bland-Altman analysis verifies that there is a somewhat negative bias between the difference of automatically extracted points and manually extracted points. Furthermore, this analysis reveals a small grouping of outlier data points. The primary cause of outliers is incorrect characterization of the number 2 as a 7, and 7 as a 2 by the OCR component.

## Conclusion

In this article, we describe a pipeline for extracting data from one of the most common figure types in biomedical literature, bar charts. Our results show that a combination of hand-coded rules and CNN algorithms can be used to detect features such as bars with high accuracy. The system's performance was tested using a collection of manually annotated PubMed articles on the subject of AMD, which focused only on VA and CFT as the primary outcomes. Based on the results, we believe that our proposed data extraction pipeline is a promising step towards a fully automated data extraction and meta-analysis system. We will add further redundancy checks in the forms of neural networks to our software in order to boost the accuracy and agreement between the manually extracted data and the output of our software. Further research on this topic will involve a larger dataset over more diverse content, including line graphs and tabulated data. This research will work toward the goal of a fully automated extraction pipeline for meta-analysis of published data.